\newcommand{\dangle}{$\cos\theta_\mathrm{CM}^{d}$}
\documentclass[final,5p,times]{elsarticle}

\usepackage{amssymb}
\usepackage{graphics}
\usepackage{graphicx}
\usepackage{xcolor}
\usepackage{sidecap}
\usepackage{microtype}
\usepackage{multirow}

\usepackage{lineno}

\journal{Physics Letters B}

\begin{document}

\begin{frontmatter}



\title{Evidence of a dibaryon spectrum in coherent  $\pi^0\pi^0 d$ photoproduction at forward deuteron angles}

\author[1]{T.C.~Jude\corref{cor1}}
\ead{jude@physik.uni-bonn.de}
\author[1]{S.~Alef\fnref{fn1}}
\author[2]{R.~Beck}
\author[4]{A.~Braghieri}
\author[5]{P.L.~Cole}
\author[1]{D.~Elsner}
\author[6]{R.~Di Salvo}
\author[6,7]{A.~Fantini}
\author[1]{O.~Freyermuth}
\author[1]{F.~Frommberger}
\author[8,9]{F.~Ghio}
\author[3]{A.~Gridnev}
\author[1]{K.~Kohl}
\author[3]{N.~Kozlenko}
\author[10]{A.~Lapik}
\author[11]{P.~Levi Sandri}
\author[10]{V.~Lisin}
\author[12,13]{G.~Mandaglio}
\author[6,11]{D.~Moricciani\fnref{fn3}}
\author[10]{V.~Nedorezov\fnref{fn3}}
\author[3]{D.~Novinskiy}
\author[4]{P.~Pedroni}
\author[10]{A.~Polonskiy}
\author[1]{B.-E.~Reitz \fnref{fn1} }%
\author[6,14]{M.~Romaniuk}
\author[1]{G.~Scheluchin\fnref{fn1}}
\author[1]{H.~Schmieden}
\author[3]{A.~Stuglev}
\author[3]{V.~Sumachev\fnref{fn3}}
\author[3]{V.~Tarakanov}

\cortext[cor1]{Corresponding author}
\fntext[fn3]{Deceased}
\fntext[fn1]{No longer employed in academia}

\address[1]{Rheinische Friedrich-Wilhelms-Universit\"at Bonn, Physikalisches Institut, Nu\ss allee 12, 53115 Bonn, Germany}
\address[2]{Rheinische Friedrich-Wilhelms-Universit\"at Bonn, Helmholtz-Institut f\"ur Strahlen- und Kernphysik, Nu\ss allee 14-16, 53115 Bonn, Germany}
\address[3]{Petersburg Nuclear Physics Institute NRC  ``Kurchatov Institute", Gatchina, Leningrad District, 188300, Russia}
\address[4]{INFN sezione di Pavia, Via Agostino Bassi, 6 - 27100 Pavia, Italy}
\address[5]{Lamar University, Department of Physics, Beaumont, Texas, 77710, USA}
\address[6]{INFN Roma ``Tor Vergata", Via della Ricerca Scientifica 1, 00133, Rome, Italy}
\address[7]{Universit\`a di Roma ``Tor Vergata'', Dipartimento di Fisica, Via della Ricerca Scientifica 1, 00133, Rome, Italy}
\address[8]{INFN sezione di Roma La Sapienza, P.le Aldo Moro 2, 00185, Rome, Italy}
\address[9]{Istituto Superiore di Sanit\`a, Viale Regina Elena 299, 00161, Rome, Italy}
\address[10]{Russian Academy of Sciences Institute for Nuclear Research, Prospekt 60-letiya Oktyabrya 7a, 117312, Moscow, Russia}
\address[11]{INFN - Laboratori Nazionali di Frascati, Via E. Fermi 54, 00044, Frascati, Italy}
\address[12]{INFN sezione Catania, 95129, Catania, Italy}
\address[13]{Universit\`a degli Studi di Messina, Dipartimento MIFT,  Via F. S. D'Alcontres 31, 98166, Messina, Italy}
\address[14]{Institute for Nuclear Research of NASU, 03028, Kyiv, Ukraine}

\begin{abstract}
	The coherent reaction, $\gamma d \rightarrow \pi^0\pi^0 d$ was studied with the BGOOD experiment at ELSA from threshold to a centre-of-mass energy of 2850\,MeV.  A full kinematic reconstruction was made, with final state deuterons identified in the forward spectrometer and $\pi^0$ decays in the central BGO Rugby Ball.  
	The strength of the differential cross section  exceeds what can be  described by models of coherent photoproduction and instead supports the three isoscalar dibaryon candidates reported by the ELPH collaboration at 2.38, 2.47 and 2.63\,GeV/c$^2$.  A low mass enhancement in the $\pi^0\pi^0$ invariant mass is also observed at the $d^*(2380)$ centre-of-mass energy which is consistent with the ABC effect.  At higher centre-of-mass energies, 
	a narrow peak in the $\pi^0 d$ invariant mass at 2114\,MeV/c$^2$ with a width of 20\,MeV/c$^2$ supports a sequential two-dibaryon decay mechanism.   
\end{abstract}



\begin{keyword}
	 BGOOD \sep dibaryon \sep coherent photoproduction

\PACS{13.60.Le,25.20.-x}

\end{keyword}

\end{frontmatter}


\section{Introduction}

Our understanding of hadron structure and relevant degrees of freedom is a crucial test for QCD in the low energy non-perturbative regime.  
Until recently, the deuteron was considered the only bound dibaryon system, however as early as 1964, Dyson and Xuong~\cite{dyson64} predicted a sextet of dibaryon states from SU(6) symmetry for baryons, denoted $\mathcal{D}_{IJ}$ for isospin $I$ and spin $J$.  With the addition of the ground state deuteron, $\mathcal{D}_{01}$, and the virtual unbound state, $\mathcal{D}_{10}$, observed as structure in $NN$ scattering, four non-strange states,  $\mathcal{D}_{12}$ and $\mathcal{D}_{21}$ ($N\Delta$), and $\mathcal{D}_{03}$ and $\mathcal{D}_{30}$ ($\Delta\Delta$) were predicted, the masses of which were determined from the deuteron mass and nucleon scattering data.  These calculations led to a plethora of searches for dibaryon states throughout the 1960s and 70s, focussing mainly on isovector ($I=1$) dibaryon candidates with mixed interpretations  (for a recent review see  ref.~\cite{clement17} and earlier reviews in refs.~\cite{locher86,strakovsky91}).  In addition to genuine dibaryon states,  One Pion Exchange models (OPE) between nucleons, for example, were also used to describe peaks observed in $\pi\pi d$ and $\pi d$ systems~\cite{yao64,barnir73}.  

    
Structures in the mass range of the suspected $d^*(2380)$ hexaquark, first identified in the fusion reaction $pn\rightarrow d\pi^0\pi^0$~\cite{adlarson11,bashkanov09} have sparked renewed interest in dibaryon searches in the non-strange sector (see ref.~\cite{clement21} for a review), particularly for isoscalar ($I=0$) dibaryon candidates.  A strong indication of the $d^*(2380)$, with $IJ^P = 03^+$, has been observed in a multitude of final states and observables~\cite{adlarson13,adlarson14,adlarson14PRC,bashkanov19,bashkanov20Pgamma,adlarson13PRC,adlarson15PLB}, and it has been considered the $\mathcal{D}_{03}$ in a $\Delta\Delta$ configuration.  This agrees with the predictions of Goldman~\cite{goldman89}, 
a three-body  $\pi N \Delta$ calculation of Gal and Garcilazo~\cite{gal14,gal13}, and is further supported by partial wave analyses (see for example refs.~\cite{adlarson14,adlarson14PRC}).  Indirect evidence of the $d^*(2380)$ may first have been observed in the 1960's via low mass enhancements of the $\pi^0\pi^0$ invariant mass~\cite{booth60,booth61,booth63} and named the ABC effect after the authors,  Abashian, Booth, and Crowe.

Models employing OPE mechanisms and ``box-diagrams" have also been used recently in an attempt to describe the $d^*(2380)$ structure~\cite{molina21,ikeno21}.  It is not clear however that such models can achieve the measured narrow width of 70\,MeV/c$^2$ and also to describe the data consistently well over the  broad range of kinematics and reaction channels that the $d^*(2380)$ has been observed in~\cite{bashkanov21}.







Numerous calculations have determined a sizeable colour confined hexaquark component of the $d^*(2380)$ (see for example, ref.~\cite{dong16}) and if it exists, its
 structure and properties may have important astrophysical implications~\cite{vidana18,bashkanov20}.  Electromagnetic production of a dibaryon directly from the deuteron ground state would impose constraints on its size and structure via transition form factors.  The ELPH collaboration measured the $\gamma d \rightarrow \pi^0\pi^0 d$ cross section~\cite{ishikawa17}  which was in agreement with the coherent photoproduction model of Fix, Arenh\"ovel and Egorov~\cite{fix05,egorov15}. 
The lowest energy data point at approximately $E_\gamma = 570$\,MeV however suggested a $d^*(2380)$ contribution, which is also supported by preliminary data from the A2 collaboration~\cite{guenther17}.  An additional ELPH dataset at  higher energies~\cite{ishikawa19} exhibited peaks at 
2.47\,GeV/c$^2$ and 2.63\,GeV/c$^2$ which could be described by the quasi-free excitation of one nucleon followed by a coalescence of the nucleons to the deuteron.  The angular distributions however were suggested to originate from two isoscalar dibaryons in the reaction mechanism.  There was also an  indication of the 2.14\,GeV/$c^2$ isovector $\mathcal{D}_{12}$ in the $\pi^0 d $ invariant mass spectrum.

The reaction $\gamma d \rightarrow \pi^0\pi^0 d$ is an ideal channel to search for dibaryons.  The isoscalar final state is only sensitive to intermediate isoscalar dibaryons, compared  to $\gamma d \rightarrow \pi^+\pi^- d$ which also has isovector coupling and  background contributions from the large $\gamma N\pi^\pm$ coupling in the Kroll-Ruderman  term (see for example ref.~\cite{fix05}).  Isovector dibaryons can still be identified in the $\pi^0 d$ invariant mass via sequential dibaryon decays.
The suppressed cross section for conventional coherent processes, which reduces quickly with momentum transfer to the deuteron, may also help to identify structure originating from dibaryon formation.

 Experimentally, a kinematic complete identification of the reaction with a clean separation of deuterons and protons is essential to separate the small number of events from coherent reactions compared to the large amount of non-coherent quasi-free events from $\gamma p(n)\rightarrow \pi^0\pi^0 p$.  The BGOOD experiment~\cite{technicalpaper} at ELSA~\cite{hillert06,hillert17} described in sec.~2 is ideal as the Forward Spectrometer cleanly separates charged particles via their mass reconstruction and the BGO Rugby Ball identifies neutral meson decays.

This letter presents differential cross section data for $\gamma p \rightarrow \pi^0\pi^0 d$ at forward deuteron angles, which cannot be described via a conventional coherent reaction as this is strongly suppressed due to the large momentum transfer to the deuteron.  The data is intended to shed more light on the dynamics of two-baryon systems at forward angles and the contrasting interpretations of earlier observed structures.  Combined with the recent measurements in the isoscalar channel referenced in this introduction, the data will hopefully aid the possible discovery and characterisation of genuine dibaryon resonances.

 


\section{Experimental setup and analysis procedure}


BGOOD~\cite{technicalpaper} is comprised of two main parts: a central region, ideal for neutral meson identification, and a forward spectrometer for charged particle identification and momentum reconstruction.  The \textit{BGO Rugby Ball} is the main detector over the central region, covering laboratory polar angles 25 to 155$^\circ$.  The detector is composed of 480 BGO crystals for the reconstruction of photon momenta via electromagnetic showers in the crystals.  The separate time readout per crystal enables a clean separation and identification of neutral meson decays.  Between the BGO Rugby Ball and the target are the \textit{Plastic Scintillating Barrel} for charged particle identification via $\Delta E-E$ techniques and the  \textit {MWPC} for charged particle tracking and vertex reconstruction.

The forward spectrometer covers a laboratory polar angle 1 to 12$^\circ$. The tracking detectors, \textit{MOMO} and \textit{SciFi} are used to track charged particles from the target.  Downstream of these is the \textit{Open Dipole Magnet} operating at an integrated field strength of 0.216\,T$\cdot$m.  A series of eight double sided \textit{Drift Chambers} track charged particle trajectories after the curvature in the magnetic field and are used to determine particle momenta with a resolution of approximately 6\,\%\footnote{The resolution improves to 3\,\% if the Open Dipole Magnet is operating at the maximum field strength of 0.432\,T$\cdot$m.}.  Three \textit{Time of Flight (ToF) Walls} downstream of the drift chambers determine particle $\beta$ and are used in combination with the measured momentum for particle identification via  mass determination.
 Track reconstruction in the Forward Spectrometer is described in ref.~\cite{technicalpaper}. 

The small intermediate region between the central region and the Forward Spectrometer is covered by \textit{SciRi}, which consists of concentric rings of plastic scintillators for charged particle detection. 

The deuterium target data presented  was taken over 26 days using an 11\,cm long target and an ELSA electron beam energy of 2.9\,GeV.
The electron beam was incident upon a thin diamond radiator\footnote{A 560\,$\mu$m thick diamond radiator mounted on a 65\,$\mu$m thick kapton foil was used to produce coherent, linearly polarised photon beam with a maximum polarisation at a beam energy of 1.4\,GeV, however the polarisation was not required for the presented analysis.} to produce an energy tagged bremsstrahlung photon beam which was subsequently collimated.  The photon beam energy, $E_\gamma$, was determined per event by momentum analysing the post bremsstrahlung electrons in the \textit{Photon Tagger}.  The total number of energy tagged photons from $E_\gamma = 450$ to 1200\,MeV was $7.5 \times 10^{12}$.  The hardware trigger used for the presented analysis required an energy tagged incident photon and an energy deposition in the BGO Rugby Ball of approximately 150\,MeV (see ref.~\cite{klambdapaper} for details).  


Candidate events were selected where exactly four photons were identified in the BGO Rugby Ball (via a veto with the Plastic Scintillating Barrel) and one charged particle in the Forward Spectrometer, corresponding to the two $\pi^0\rightarrow \gamma\gamma$ decays and a forward going deuteron.  Events were rejected if any additional charged particle was identified.

The invariant mass of each two photon system  was required to be within 40\,MeV of the $\pi^{0}$ mass (corresponding to approximately 2.5$\sigma$).  For a given event, all combinations where two $\pi^0$ from the four photons could be reconstructed were retained for further analysis. 

Two mass determinations of the forward going particle were made per event.  The first was the mass reconstruction in the forward spectrometer via momentum and $\beta$ measurements, referred to as the \textit{measured mass}.  The second was the missing mass recoiling from the $\pi^0\pi^0$ system, referred to as the \textit{$\pi^0\pi^0$ missing mass}.  For the dominant quasi-free  background reaction, $\gamma p(n)\rightarrow \pi^0\pi^0 p$, the \textit{measured mass} is peaked at the proton mass, whereas the \textit{$\pi^0\pi^0$ missing mass} is higher than the deuteron mass and increases with $E_{\gamma}$ due to the assignment of the target mass to that of the deuteron.

\begin{figure} [htb]
 	\centering
 	\vspace*{0cm}
 	\resizebox{\columnwidth}{!}{%
 		\includegraphics{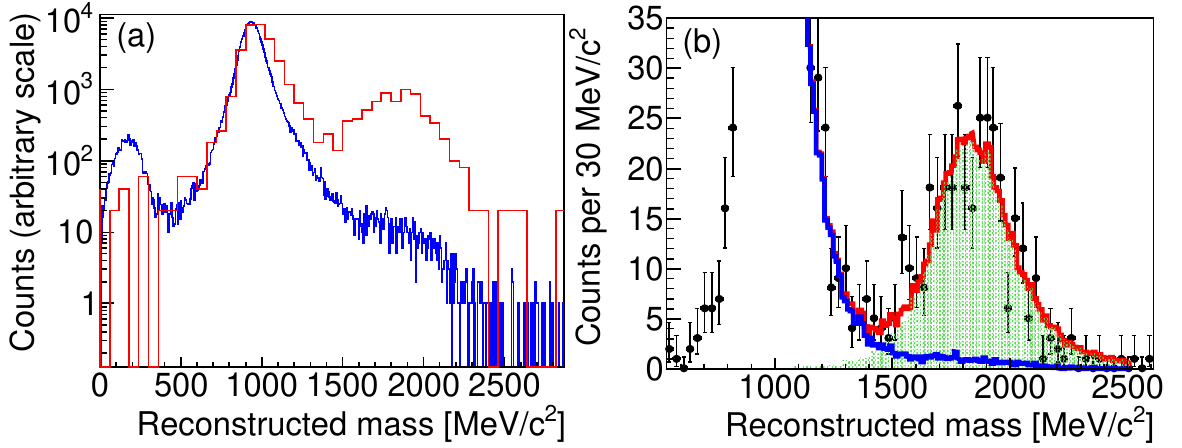}
 	}
 	\caption{\textit{Measured mass}, with the requirement that 2$\pi^0$ are identified in the BGO Rugby Ball and the forward going particle momentum is between 800 to 950\,MeV/c. (a)  All events from the dataset are shown in blue.  The red line corresponds to events passing additional selection criteria described in the text and scaled by a factor of 2000.  (b)  A fit to the events (black circles) passing the additional selection criteria, including simulated $\pi^0\pi^0d$ events (shaded green area) and quasi-free proton background (blue curve).  The total fit is the red curve.}		
 	\label{fig:massfit}
 \end{figure}

 The blue histogram in Fig.\ref{fig:massfit}(a) shows the \textit{measured mass} after two $\pi^{0}$ have been identified.  There is a small peak at the $\pi^+$ and a larger peak at the proton mass from quasi-free production.  The peak corresponding to deuterons is barely discernible in the high energy tail of the proton candidates.   Additional selection criteria were used to enhance the signal from deuterons in order to visualise the mass distribution.  The angle between the missing momentum from the $\pi^0\pi^0$ system and the forward going charged particle was required to be smaller than 7.5$^\circ$ and the \textit{$\pi^0\pi^0$ missing mass} was required to be lower than 1900\,MeV/c$^2$ to remove background from quasi-free production off the proton.
 The red histogram in Fig.\ref{fig:massfit}(a)  (scaled to equal the number of protons in the blue spectrum) shows the \textit{measured mass} after these selection criteria, where a peak at the deuteron mass is clear.   Fig.\ref{fig:massfit}(b) shows the same data (not scaled and with finer binning), where a fit has been made using simulated $\gamma d \rightarrow \pi^0\pi^0 d$ events and real data of quasi-free production off the proton, with good agreement to the data. 

The mass distribution in Fig.\ref{fig:massfit}(b) was used to determine a sensible deuteron mass selection cut for the \textit{measured mass}, after which the selection criterion on the \textit{$2\pi^0$ missing mass} to enhance the deuteron signal was removed.  The \textit{measured mass}  was subsequently required to be between 1550 to 2500\,MeV/c$^2$, corresponding to $1.3\sigma$ below to $2.6\sigma$ above the deuteron mass (approximating the distribution as Gaussian).  The asymmetric selection limited the contribution of background from quasi-free protons. 

\begin{figure} [htb]
	\centering
	\vspace*{0cm}
	\resizebox{\columnwidth}{!}{%
		\includegraphics[trim={1.0cm 2.2cm 0.4cm 2.1cm},clip]{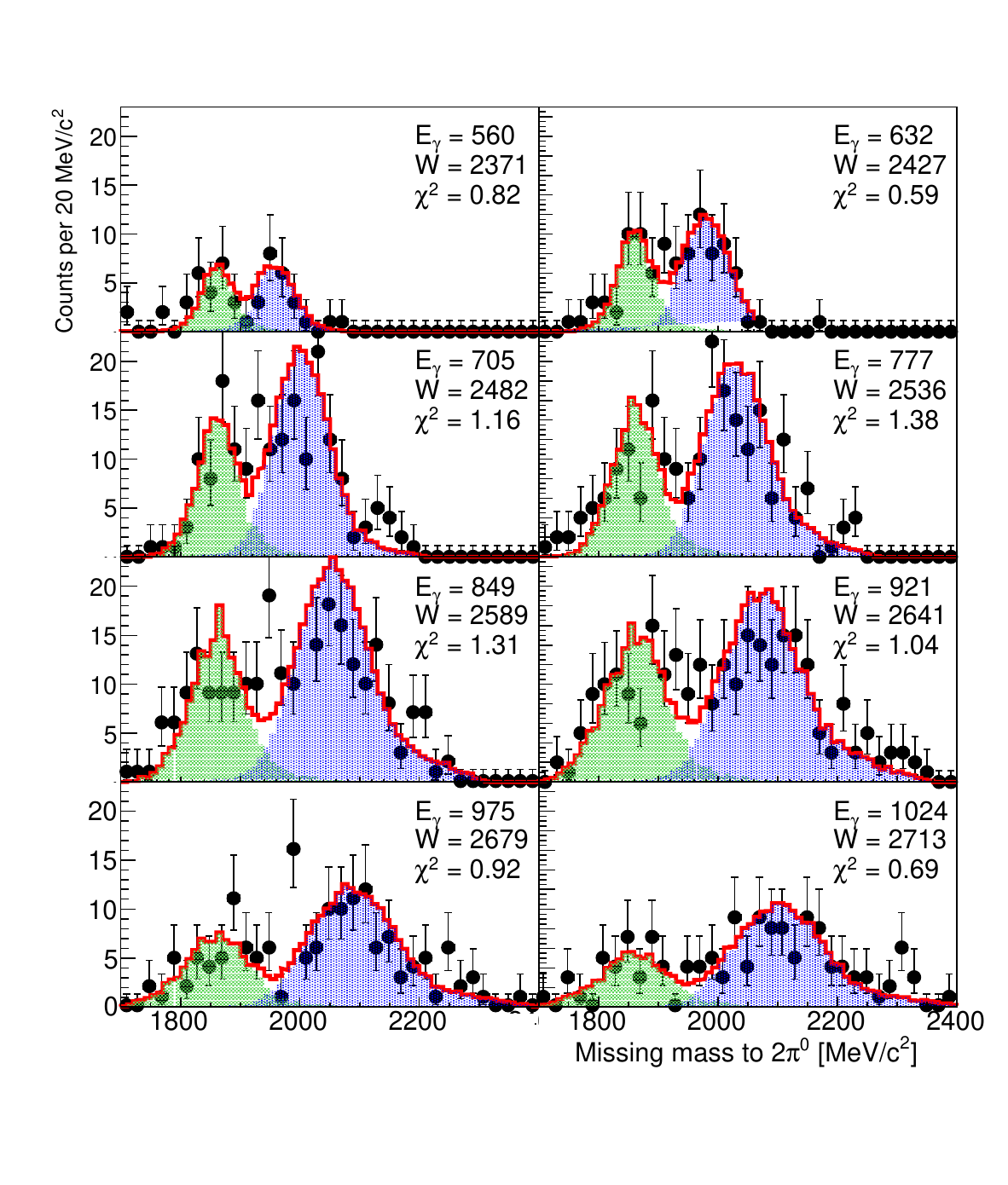}
	}
	\caption{\textit{$\pi^0\pi^0$ missing mass} after the selection criteria described in the text.  Every other photon beam energy interval, $E_{\gamma}$ is shown up to 1024\,MeV, and labelled inset in units of MeV together with $W$ and the $\chi^2$ per degree of freedom of the fit.  The data are the black circles, and the fit to the data is the thick red line.  Fitted contributions from simulated $ \pi^0\pi^0 d$ and background from quasi-free events off the proton are shown as the shaded green and blue spectra respectively.}		
	\label{fig:mmfit}
\end{figure}

The \textit{$\pi^0\pi^0$ missing mass} after these selection criteria is shown in Fig.~\ref{fig:mmfit}.  A peak at the deuteron mass of 1876\,MeV/$c^2$ is evident, with background from quasi-free production off the proton at higher mass.  This background contribution is described by changing the \textit{measured mass} selection to between 600 to 1300 MeV/$c^2$ over the proton mass\footnote{An alternative method to describe the quasi-free background was studied using liquid hydrogen data and selecting forward particles with a \textit{measured mass} over the deuteron mass region.  The lower statistics and the absence of the target Fermi motion however demonstrated that selecting the quasi-free events from the liquid deuterium data was preferred.}.
 A fit was made to extract the yield of $\gamma d \rightarrow \pi^0\pi^0 d$ events  using this quasi-free background and simulated $\pi^0\pi^0 d$ events with a phase space distribution.  The two distributions, shown in fig.~\ref{fig:mmfit} as the shaded green and blue regions are scaled to minimise the $\chi^2$ per degree of freedom for each $E_\gamma$ interval.


\begin{figure} [htb]
	\centering
	\includegraphics[width=\columnwidth,trim={0cm 0cm 0 0cm},clip]{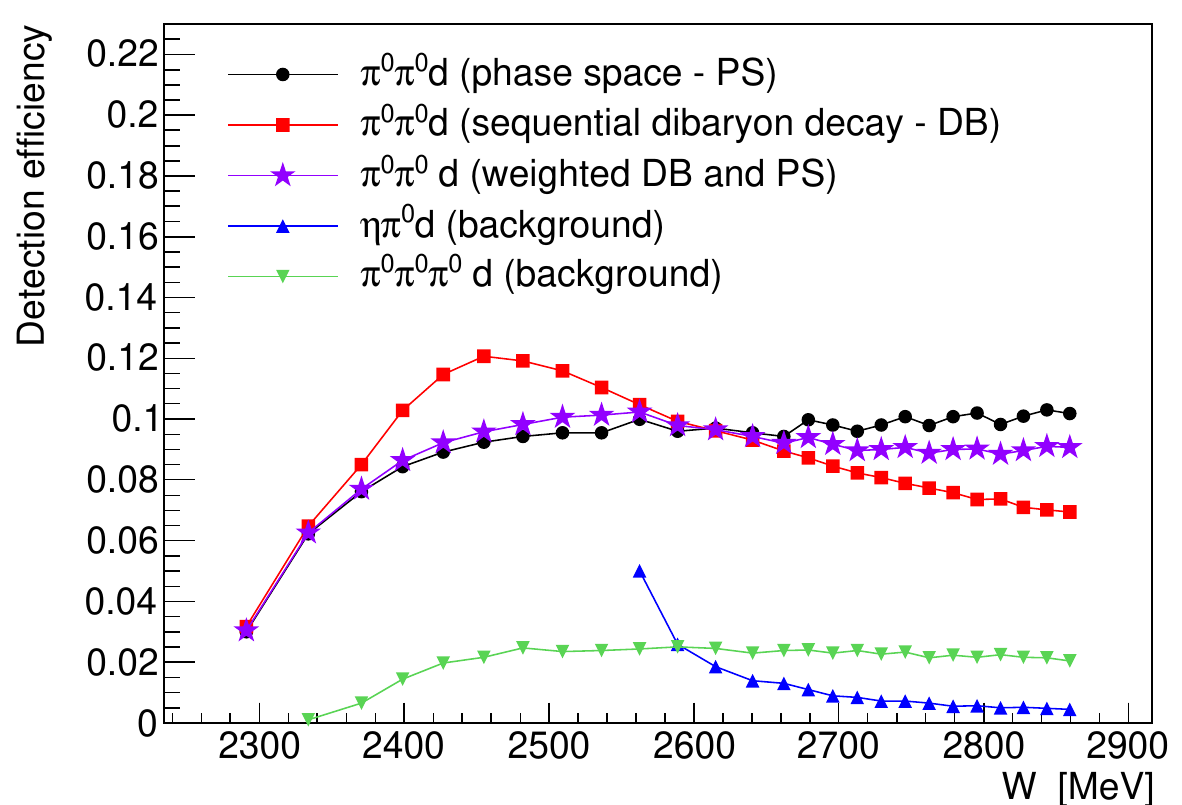}
	\caption{$\gamma d \rightarrow \pi^0\pi^0 d$ detection efficiency for \dangle{} $> 0.8$ assuming a phase space distribution and an intermediate isovector dibaryon  (black circles and red squares respectively) .  The efficiency of detecting background channels despite selection criteria removing most events are shown as the green and blue triangles for $\gamma d \rightarrow \pi^{0}\pi^0\pi^0 d$ and $\gamma d \rightarrow \eta\pi^0 d$ respectively.}
	\label{fig:deteff}
\end{figure}

Fig.~\ref{fig:deteff} shows the detection efficiency for the reconstruction of $\gamma d \rightarrow \pi^0\pi^0 d$ events assuming a phase space distribution  and if a sequential decay of two dibaryons is assumed (discussed later in sec. 3).  
The differential cross section in sec.~3 used the weighted detection efficiency of these two distributions, indicated by the purple stars.


\begin{figure} [h]
	\centering
	\vspace*{0cm}
	\resizebox{\columnwidth}{!}{%
		\includegraphics[trim={0cm 0cm 2cm 2.0cm},clip]{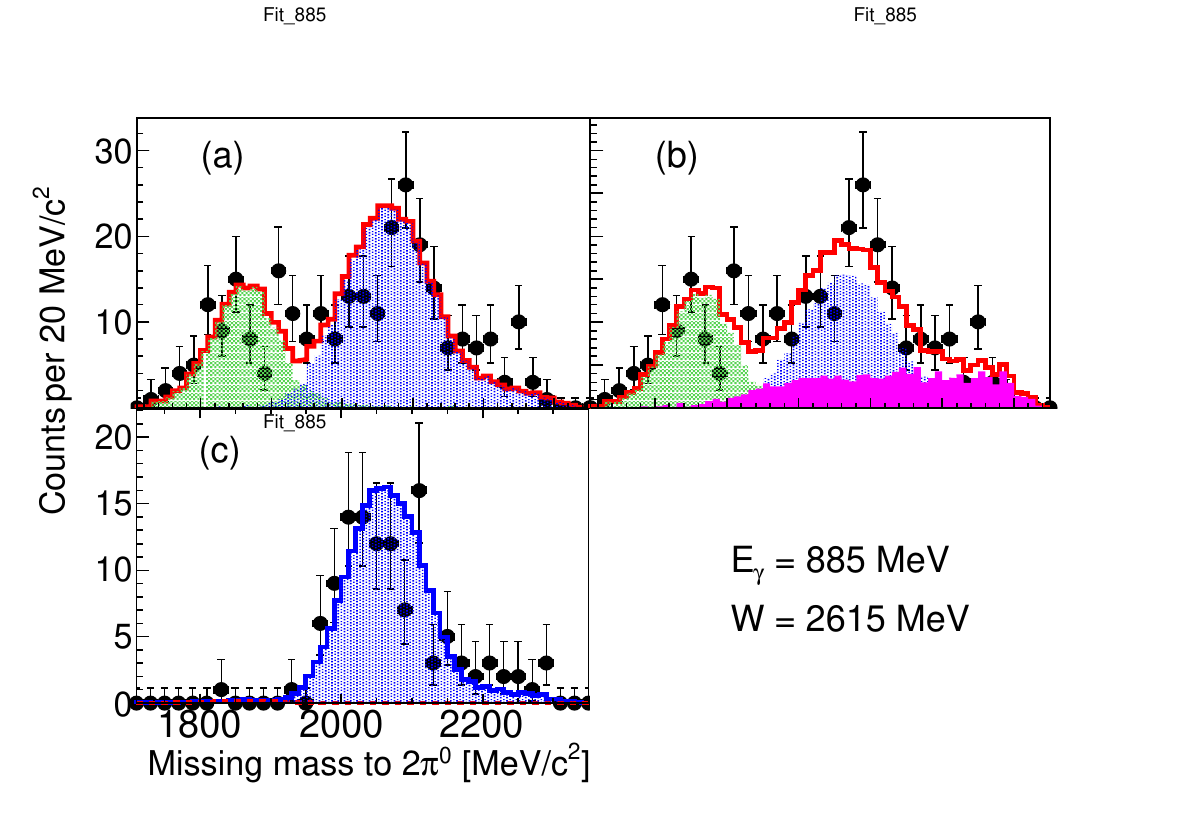}
	}
	\caption{An example of systematic uncertainties evident in $\pi^0\pi^0$ \textit{missing mass} at $E_{\gamma} = 885$\,MeV.  (a)  A ``standard" fit using the simulated signal and quasi-free proton background, as is shown in fig.~\ref{fig:mmfit}.  (b)  The inclusion of additional background from simulated $\gamma d \rightarrow 3\pi^{0} d$ events (filled magenta).  (c)  Using a liquid hydrogen target.}
	\label{fig:MMFit_SysError}
\end{figure}

Simulated coherent production of $3\pi^0 d$ and $\eta\pi^0 d$ events were used to investigate possible background from these channels where the cross sections are poorly known.  The false efficiency of the extent that these channels pass the selection criteria is also shown in fig.~\ref{fig:deteff}.  With approximately 25\,\%   relative to the $\pi^0\pi^0 d$ efficiency, it is clear that they cannot be neglected. 
The $\pi^0\pi^0$ missing mass spectra were therefore also fitted including the additional simulated background channels.
The fit could accommodate a significant fraction of $3\pi^0 d$ background, an example of which is shown in fig.~\ref{fig:MMFit_SysError}(b).  This additional background however was predominantly under the quasi-free proton events and did not significantly affect the extracted yield of the $\pi^0\pi^0 d$ channel or the $\chi^2$ per degree of freedom of the fit over most of the beam energy range.  The systematic uncertainty in the yield extraction arising from this background contribution increased approximately exponentially with beam energy, and was estimated as 5\,\%, 12\,\% and 26\,\% for $E_\gamma$ =  900, 1000 and 1100\,MeV respectively.  The systematic uncertainty arising from reactions off the target cell windows was determined by an equivalent analysis of data using a hydrogen target.  Shown in Fig.~\ref{fig:MMFit_SysError}(c), it is clear there is negligible signal with only background from reactions off the proton.  The uncertainty was therefore considered to be 1\,\% or less.

 Systematic uncertainties from the modelling of the detector efficiencies (3.6\,\%), forward track finding  (1.0\,\%), timing cuts (2.0\,\%) and beam spot alignment  (4.0\,\%) were measured previously~\cite{klambdapaper}.  Systematic uncertainties from the $\pi^0\rightarrow \gamma\gamma$ identification (3.5\,\% for both $\pi^0$) and the angle selection cut between the missing and measured deuteron momentum (1.0\,\%) were determined by an equivalent analysis of the reaction $\gamma p \rightarrow \pi^0\pi^0 p$ with a hydrogen target where the beam and trigger conditions were identical to the deuterium target dataset. 
 The measured differential cross section (not shown) gave a good agreement to previous results from the A2 Collaboration~\cite{kashevarov12}.
 
 
\section{Results and interpretation}

\subsection{The  $\gamma d \rightarrow \pi^0\pi^0 d$ differential cross section versus $W$}
\begin{figure} [htb]
	\centering
	\vspace*{0cm}
	\resizebox{\columnwidth}{!}{%
		\includegraphics{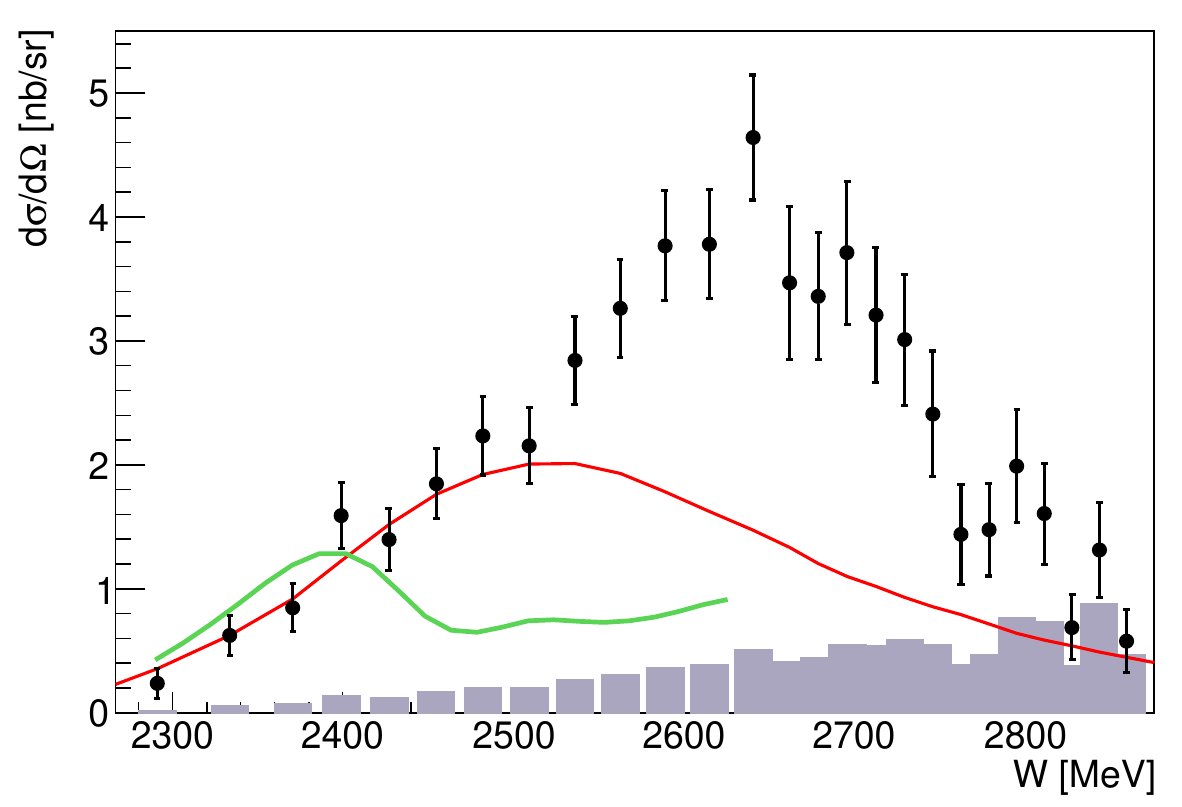}
		
	}
	\caption{$\gamma d \rightarrow \pi^{0}\pi^{0} d $ differential cross section for \dangle$ > 0.8$.  Systematic errors are the grey bars on the abscissa.  Superimposed is the model prediction from Fix, Arenh\"ovel and Egorov ~\cite{fix05,egorov15} scaled by a factor of five (green line), and the toy pickup model set at an arbitrary scale (red line, see text for details).}
	\label{fig:cscomplete}
\end{figure}

The differential cross section for \dangle{} $> 0.8$ (where $\theta^d_\mathrm{CM}$ is the centre-of-mass polar angle of the deuteron) is shown in fig.~\ref{fig:cscomplete}.  The data peaks at  $W \sim 2650$\,MeV with a cross section of 4\,nb/sr.  This is approximately an order of magnitude higher than the model prediction of Fix, Arenh\"ovel and Egorov~\cite{fix05,egorov15} which assumed coherent production off the deuteron, where at forward angles the cross section falls very quickly due to the increasing momentum transfer.  For \dangle{} $> 0.8$ at $W = 2300$ and 2800\,MeV, the three-momentum transfer to the deuteron is 0.4 and 1.0\,GeV/c respectively, which is much higher than the Fermi momentum of the constituent nucleons (typically 80\,MeV/c) and therefore what can be transferred to the deuteron for it to remain intact. 

\begin{figure} [htb]
	\centering
	\includegraphics[width=\columnwidth,trim={0cm 20.5cm 1.5cm 0cm},clip]{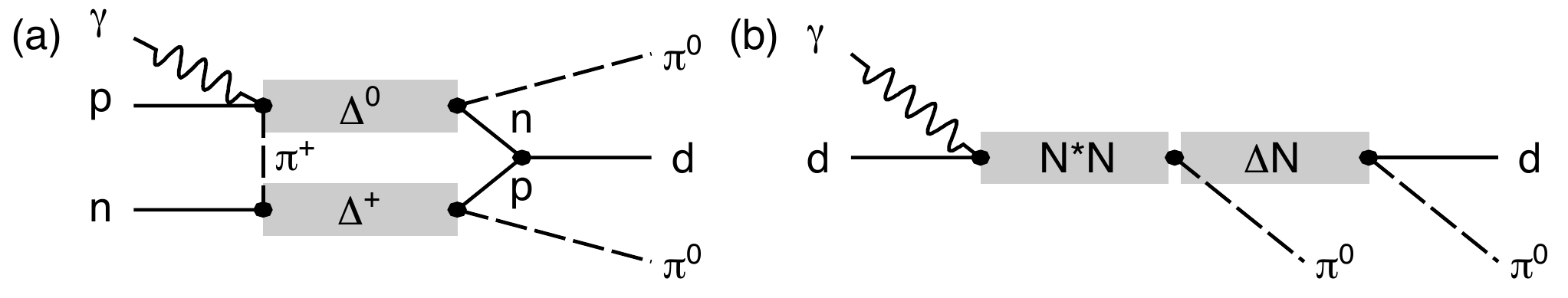}
	\caption{Possible mechanisms contributing to the $\gamma d \rightarrow \pi^0\pi^0 d$ reaction.  (a)  The \textit{Toy pickup model} described in the text.  (b)  A sequential dibaryon decay mechanism.}		
	\label{fig:seqdecay}
\end{figure}

Fig.~\ref{fig:cscomplete} also shows the results from a {\it toy pickup model} depicted in fig.~\ref{fig:seqdecay}(a)  arbitrarily normalised to fit the data.  The model does not include intermediate dibaryons, but instead  $\Delta^{0} \pi^+$ are produced in a quasi-free reaction off the proton  and the $\pi^{+}$ subsequently rescatters off the spectator neutron, producing a $\Delta^+$.  This model uses the same diagram as a conventional $\Delta\Delta$ excitation by $t$-channel meson exchange  that was previously suggested as a mechanism for the ABC effect~\cite{risser73,bar75,alvarez98,alvarez99} before it was attributed to the  $d^*(2380)$.  In the model, the proton and neutron from the decays of the $\Delta$s coalesce to form the deuteron in the final state if their relative momentum is smaller than a momentum sampled from the  internal Fermi momentum distribution of the deuteron. On-shell momentum and energy conservation is assumed, with the  Breit-Wigner mass and width of the $\Delta$ and the spin dependent $(1+3\cos^2\theta)$ decay angular distribution.  A uniform initial state photon energy distribution was used, and the output was weighted according to the pion exchange propagator, $1/(m_\pi^2 + q^2)^2$,  where $q$ is the pion momentum in the centre-of-mass frame of the $\pi^+$ $n_\mathrm{spectator}$ system.  An additional $q^2$ weighting was assumed for the dominant magnetic coupling at the $\gamma p \Delta^0\pi^+$ vertex.  The model is simplistic and used only to judge if this or a similar mechanism could provide a significant contribution to the observed spectrum and if so over which range of $W$. The toy model produces a broad distribution which peaks roughly at twice the $\Delta$ mass.  The distribution is at an arbitrary scale and would need to be weighted by the $\Delta\pi$ photoproduction cross section for an accurate determination, however it is clear that the width and shape cannot describe the $\gamma d \rightarrow \pi^0\pi^0 d$ excitation spectrum alone.
Other combinations of higher lying resonances were also investigated, where the mass distributions became broader and moved to higher energies, but also could not describe the data. 


\begin{figure} [htb]
	\centering
	\vspace*{0cm}
	\resizebox{\columnwidth}{!}{%
		\includegraphics{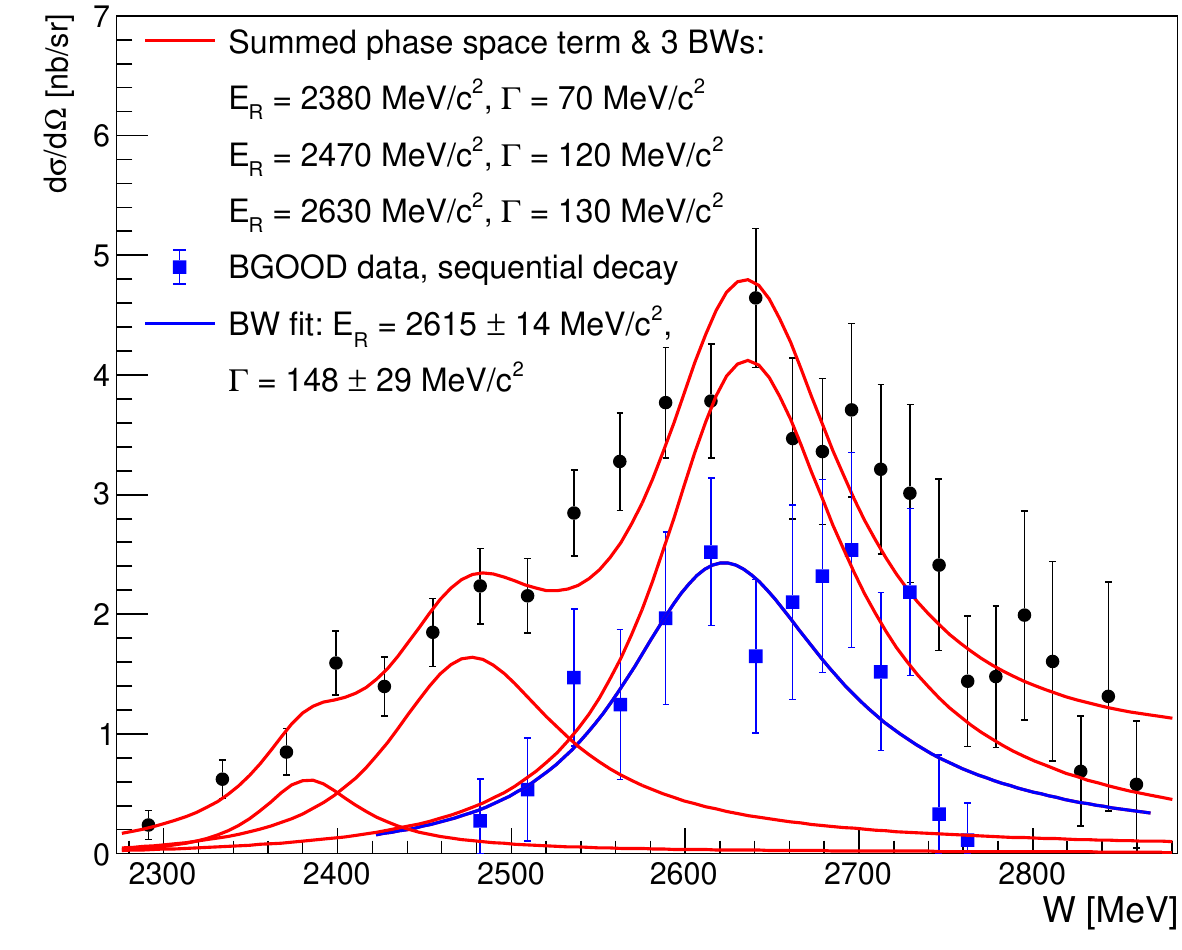}
		
	}
	\caption{$\gamma d \rightarrow \pi^{0}\pi^{0} d $ differential cross section for \dangle$ > 0.8$ (the same as in fig.~\ref{fig:cscomplete}).  A fit including three Breit-Wigner functions (BW) are shown as the red lines, with the fixed masses and widths labelled inset.  The additional small centre-of-mass momentum term described in the text is not shown.  The blue square data points are the differential cross section for the first of the sequential dibaryon candidate determined from the $\pi^0 d$ invariant mass distributions shown in fig.~\ref{fig:pidmass}  
	with a Breit-Wigner function fitted and the mass and width labelled inset.
	Only the statistical uncertainties are shown.}
	\label{fig:cscompleteBW}
\end{figure}

Fig.~\ref{fig:cscompleteBW} shows the same differential cross section data but fitted with the $d^*(2380)$ and the two additional isoscalar dibaryons reported by the ELPH collaboration~\cite{ishikawa19}.  A Breit-Wigner function is assumed for each dibaryon candidate, with masses and widths fixed from ref.~\cite{ishikawa19} but not their relative amplitudes.  A term proportional to the square of the centre-of-mass momentum is also included.  This is a small contribution, rising slowly with $W$ and approximates the phase space available to the deuteron and two $\pi^0$ in the final state.  
The fit used the statistical and fitting systematic uncertainties summed in quadrature and achieved a $\chi^2$ per degree of freedom of 0.96.  The fit therefore is consistent with the three dibaryon scenario, however the limited statistical precision and resolution in $W$ cannot rule out other reaction mechanisms.  The inclusion of the $d^*(2380)$ is tentative, for example, as a reasonable $\chi^2$ per degree of freedom of 1.35 is achieved if it is omitted and any small enhancement at low $W$ could also be described by  variations of the toy pickup model described above.  

The blue squares and fit showing a proposed sequential dibaryon decay are discussed in sec.~3.3.

\subsection{$\pi^0\pi^0$ and $\pi^0d$ invariant mass distributions over the $d^*(2380)$ mass range}

Fig.~\ref{fig:pidmass1} shows the invariant mass of the $\pi^0 d$ and $\pi^0 \pi^0$ systems for $W$ from 2270 to 2441\,MeV (corresponding to five tagger channels).  This is centred over the $d^*(2380)$ mass, covering approximately 90\,\% of the Breit-Wigner function in fig.~\ref{fig:cscompleteBW}.   Background from quasi-free production off the proton was subtracted by scaling the background according to the fit to the \textit{$\pi^0\pi^0$ missing mass} spectra in fig.~\ref{fig:mmfit}.  Only events within $3\sigma$ of the deuteron mass in the missing mass spectra were included to reduce the amount of quasi-free background required to be subtracted.  An additional systematic uncertainty of 10\,\% for the subtraction of this background was estimated by varying the selection cut, giving a total of 13.5\,\% when combined in quadrature with the other systematic uncertainties.  

\begin{figure} [h]
	\centering
	\vspace*{0cm}
	\resizebox{\columnwidth}{!}{%
	\includegraphics{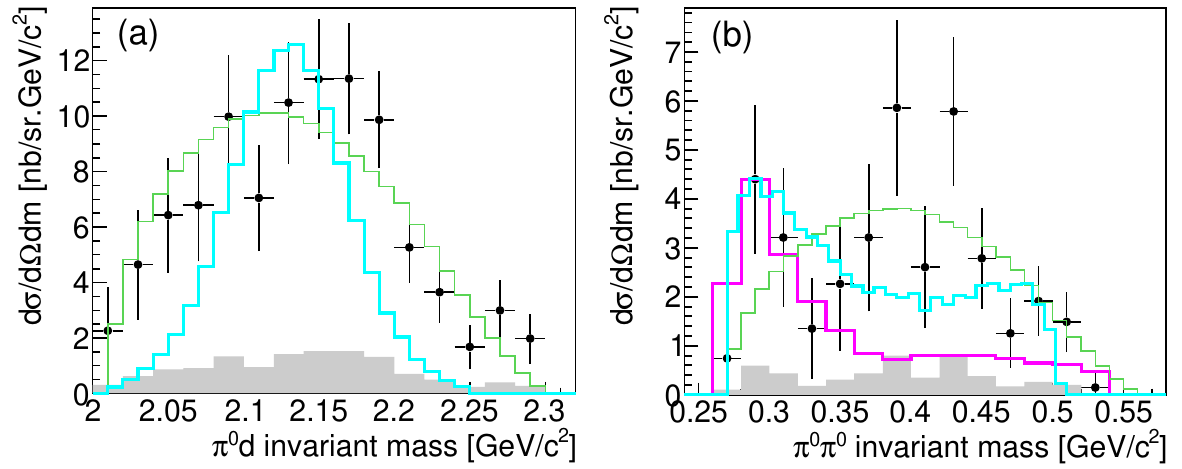}
	}
	\caption{Differential cross section versus the invariant mass of the (a) $\pi^{0}d$ and (b) $\pi^0\pi^0$ system for \dangle{} $>0.8$ and $W$ from  2270 to 2441\,MeV.  The measured data are the black data points and the systematic uncertainties are the grey bars on the absisca.  The green line is the phase space distribution with an integral equal to the measured data.   Two distributions from the ABC effect are shown at an arbitrary scale, from double pionic fusion to $^4$He~\cite{adlarson12} (cyan lines) or deuterium~\cite{adlarson11} (magenta line, only shown in (b)).}
	\label{fig:pidmass1}
\end{figure}

The phase space distribution gives a reasonable description of the $\pi^0d$ invariant mass in fig.~\ref{fig:pidmass1}(a), with qualitative agreement to the $\pi^0\pi^0$ invariant mass in the previous photoproduction data from the ELPH collaboration~\cite{ishikawa17}.  Also shown is the invariant mass for double pionic fusion to deuterium at $W= 2380$\,MeV~\cite{adlarson11}, which is narrower than the data. 

The $\pi^0\pi^0$ invariant mass in fig.~\ref{fig:pidmass1}(b) exhibits a low mass enhancement and a dip at approximately 0.34\,GeV/c$^2$.  This also appears to be in agreement with the ELPH collaboration data and is similar to the ABC effect in refs.~\cite{adlarson11,bashkanov09} which was attributed to the $d^*(2380)$.  The cyan and magenta lines (scaled to match the data at  0.29\,GeV/c$^2$) show the  distribution in double pionic fusion to deuterium~\cite{adlarson11} and  $^4$He~\cite{adlarson12}  respectively, where in both cases it is interpreted that an intermediate $d^*(2380)$ is formed.  Qualitatively there is a good agreement and is preferred over the phase space spectrum, despite limited statistical precision.  A fit was made (not shown) to the data including the double pionic fusion to deuterium spectrum  and a Gaussian distribution centred around 0.4\,GeV/c$^2$ to describe the higher invariant mass range.  A $\chi^2$ per degree of freedom of 1.18 was achieved, which is an improvement over 1.39 when fitting with only a phase space distribution. A differential cross section for $\gamma d \rightarrow d^*(2380) \rightarrow \pi^0\pi^0 d$ was subsequently determined as $(22 \pm 6_\mathrm{stat} \pm 4_\mathrm{sys})$\,nb/sr for  \dangle $> 0.8$.  The systematic uncertainty was estimated by varying fit parameters and by comparing the fits using the different double pionic fusion spectra.  The angular distribution of the $d^*(2380)$ is already well determined in fusion reactions (see for example, ref.~\cite{adlarson11}), and so the total cross section can be extrapolated to $(11.3 \pm 3.2_\mathrm{stat} \pm 2.7_\mathrm{sys})$\,nb. 
This is of a similar order of magnitude to what was determined using an extended chiral constituent quark model in ref.~\cite{dong19}, where the resonance peak was calculated (assuming certain caveats) as 1.4\,nb.  With improved statistics, this fitting method to the ABC effect can enable a particularly accurate $d^*(2380)$ photoproduction cross section measurement.


\subsection{Proposed sequential dibaryon decay for $W > 2500$\,MeV}

For $W$ higher than 2500\,MeV, a double peaking structure is observed in the $\pi^0 d$ invariant mass (fig.~\ref{fig:pidmass} (left panels)),which is similar to what was observed by the ELPH Collaboration~\cite{ishikawa19}, where it was interpreted as an isovector dibaryon with a mass of $2140 \pm 11$\,MeV/c$^2$ and a width of $91\pm 11$\,MeV/c$^2$ from the  decay of an isoscalar dibaryon.
This is depicted in Fig.~\ref{fig:seqdecay}(b), where the $N^{*}N$ and $\Delta N$ configurations are proposed as dibaryons in the sequential decay.  This reaction mechanism was input to the BGOOD simulation where the mass and width of the isoscalar dibaryon was varied until a comparison to the real data over multiple $W$ intervals achieved a minimal $\chi^2$.  Shown in fig.~\ref{fig:pidmass} (left panels) as the blue line with an additional phase space contribution in green,  a mass of 2117\,MeV/c$^2$ and a width of 20\,MeV/c$^2$ proved optimal.   The higher energy broader peak is the reflection of the uncorrelated $\pi^0 d$ combination.  A systematic uncertainty of 10\,MeV/c$^2$ is attributed to the mass due to the energy calibration of the BGO Rugby Ball.   The mass is comparable to the mass of 2140\,MeV/c$^2$ determined by the ELPH collaboration when accounting for the statistical and systematic uncertainties in both measurements.

The measured 20\,MeV/c$^2$ width of the dibaryon candidate is approximately the same as the experimental $\pi^0 d$ mass resolution and can therefore be considered an upper limit.  This is much narrower than the width of 91\,MeV/c$^2$ reported by the ELPH collaboration, which cannot be accounted for in this data and cannot give a satisfactory fit.   The fact that the peak is below the $N\Delta$ threshold and is much narrower than the $\Delta$ may suggest this is a genuine dibaryon state.   
Similarly to the ELPH measurement however, a phase distribution for each of the decays was assumed, which does not take into account  angular momentum and a properly coupled two $\pi^0$ system, the dynamics of which may affect the $\pi^0 d$ invariant mass distributions.  These reasons and alternative mechanisms may yet explain the small deviation in mass and significantly different width of the observed structure. 
 
  

\begin{figure} [h]
	\centering
	\vspace*{0cm}
	\resizebox{\columnwidth}{!}{%
		\includegraphics{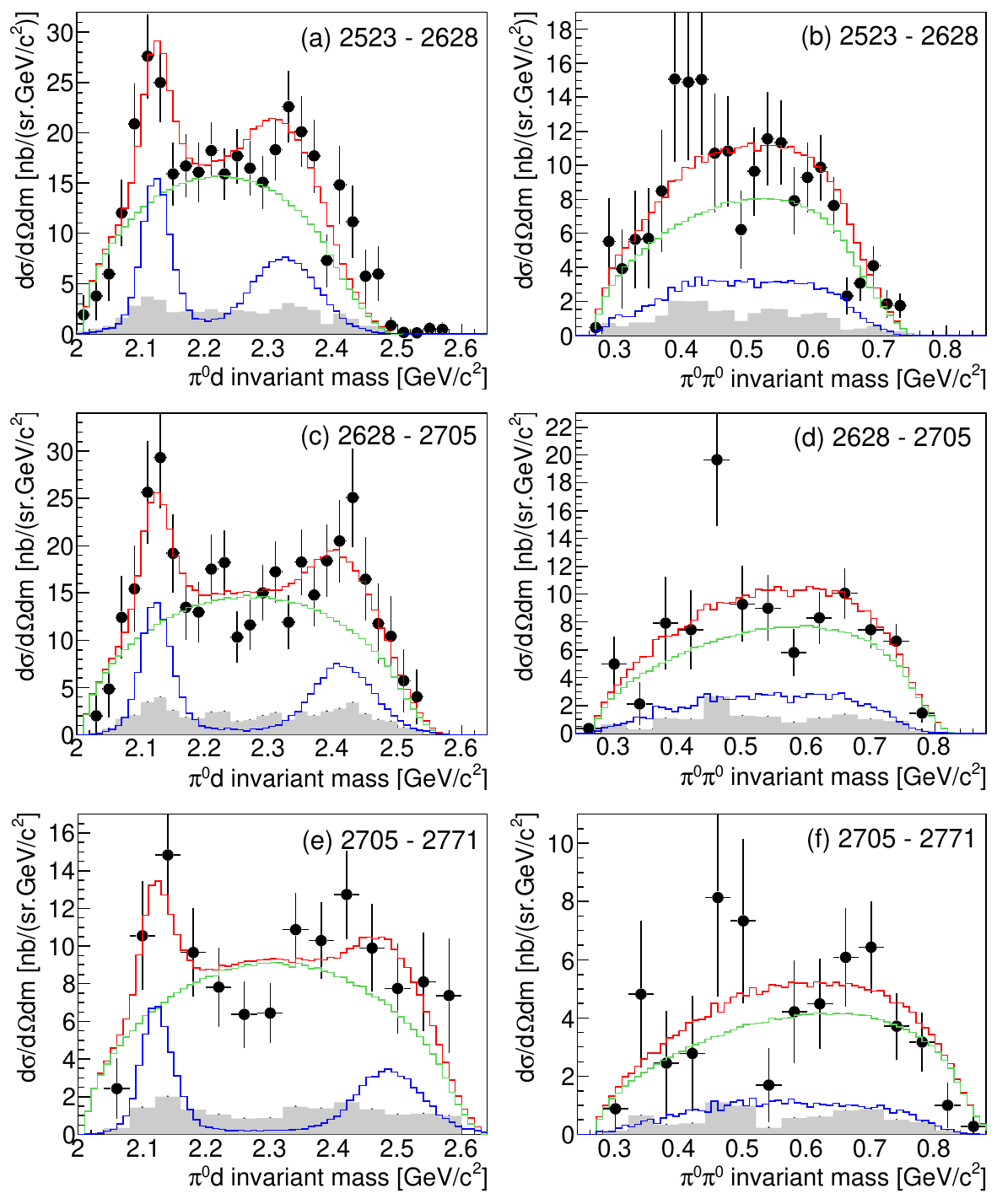}
	}
	\caption{Differential cross section versus the invariant mass of the (left panels) $\pi^{0}d$ and (right panels) $\pi^0\pi^0$ system for \dangle{} $>0.8$ and $W$ intervals labelled inset (corresponding to four tagger channels each).  The measured data are the black data points and the systematic uncertainties are the grey bars on the absisca.   The fitted distribution to the $\pi^0d$ invariant mass (red line) is comprised of phase space (green line) and proposed sequential dibaryon decay (blue line) contributions.}		
	\label{fig:pidmass}
\end{figure}

The proposed dibaryon sequential decay and phase space contributions are superimposed onto the $\pi^0\pi^0$ invariant mass distributions in fig.~\ref{fig:pidmass} (right panels).  Unsurprisingly, the two distributions are similar as the simulation used relative $s$-wave (phase space) distributions for the dibaryon decays.  Whilst this distribution describes the data reasonably well, in all three $W$ intervals there appears some structure between 0.4 to 0.5 GeV/c$^2$ which may hint at dynamics beyond relative $s$-wave decays, albeit limited by statistical precision.
This structure in the $\pi^0\pi^0$ invariant mass distribution can be more clearly seen in fig.~\ref{fig:dalitz}, where $W$ extends over the full range of the sequential dibaryon decay candidate.   A dip at 0.45\,GeV/c$^2$ is observed, where the strength reduces by approximately 50\,\%.  The modelled distribution gives a poor description with a $\chi^2$ per degree of freedom of 2.55, however an improvement can be made by including dibaryon decays with higher angular momenta.  Shown in  fig.~\ref{fig:dalitz}, changing the angular momentum of both the first and second dibaryon decays to relative $p$-wave for example gives an improved fit with a $\chi^2$ per degree of freedom of 1.78.  This is only a demonstration of the impact the angular distributions may have, without including dynamics which may arise from a properly coupled two $\pi^0$ system.


\begin{figure} [h]
	\centering
		\includegraphics[trim={0cm 0cm 0cm 0cm},clip,width=0.7\columnwidth]{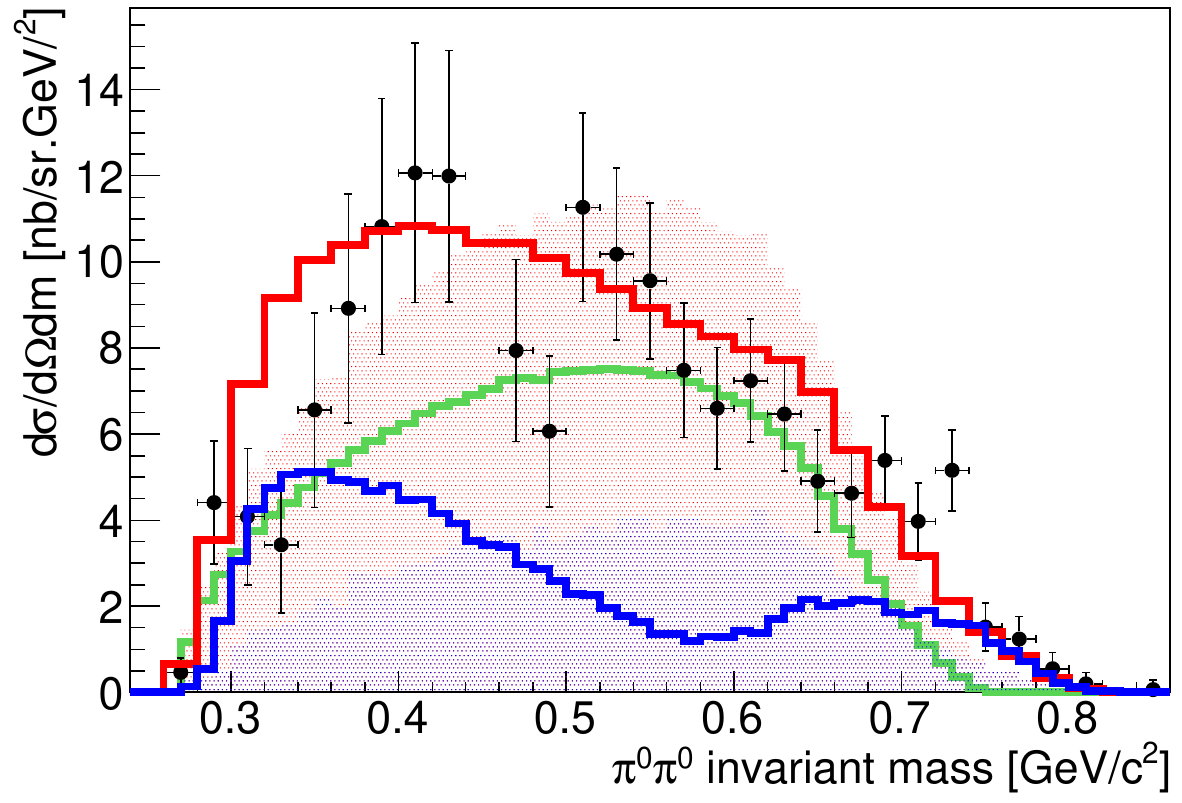}
	\caption{$\pi^0\pi^0$ invariant mass distribution for 2523 $< W < 2738$\,MeV (over the proposed sequential dibaryon decay) .  The shaded red distribution is comprised of the phase space (thick green line) and two sequential dibaryon decays both in $s$-wave (shaded blue area).
		The thick red line is comprised of the phase space and the 
		two sequential dibaryon decays both in $p$-wave (thick blue line).}		
	\label{fig:dalitz}
\end{figure}


The $\pi^0 d$ invariant mass for each $W$ interval was fitted  to extract the differential cross section of the proposed sequential decay in fig.~\ref{fig:seqdecay}(b), shown as the blue squares in fig.~\ref{fig:cscompleteBW}.  
A Breit-Wigner function was fitted, extracting a mass of ($2618 \pm 14$)\,MeV/c$^{2}$ and width of ($148 \pm 29$)\,MeV/c$^{2}$ with a $\chi^2$ per degree of freedom of 1.18.  Both the mass and width agree with the proposed highest mass dibaryon from the ELPH collaboration~\cite{ishikawa19} (also fitted in fig.~\ref{fig:cscompleteBW}).  
A full implementation of angular momentum dynamics is required to properly describe any sequential decay, however the evidence is at least consistent with the observation of two decay modes of an isoscalar dibaryon, either directly to $\pi^0\pi^0 d$ or to $\pi^0 \mathcal{D}_{12}$, where the $\mathcal{D}_{12}$ is an $N\Delta$ configuration.  It is interesting to note that the mass and width are compatible with a bound $N(1680)5/2^+N$ dibaryon system.  Numerous other \textit{four star} resonances\footnote{As defined by the Particle Data Group~\cite{pdg20}.} have a similar mass to the $N(1680)$, however  couple less strongly to $\pi N$ and with smaller photoproduction cross sections.  The positive parity also requires a decay with odd relative  angular momentum to the $N\Delta$ $\pi^0$ system, which is both supported by the $\pi^0\pi^0$ invariant mass distribution, where two relative $p$-wave decays give an improved description, and by the change in spin required of the constituents in the two dibaryon candidates. 
 
\section{Conclusions}

The differential cross section for the coherent reaction, $\gamma d \rightarrow \pi^0\pi^0 d$ has been measured at BGOOD for \dangle$ > 0.8$ via the identification of $\pi^0\rightarrow \gamma\gamma$ in the BGO Rugby Ball and the deuteron in the Forward Spectrometer.  In this kinematic regime, the momentum transferred to the deuteron is significantly higher than the internal Fermi momentum and prevents the strength and shape of the excitation spectra being described by the conventional coherent photoproduction models of Fix, Arenh\"ovel and Egorov~\cite{fix05,egorov15}, or by a simplistic toy model of quasi-free production where final state nucleons coalesce to form the deuteron via $\Delta\pi$ rescattering.  Instead, the data supports the findings of the ELPH collaboration, with proposed isoscalar dibaryon states at 2380, 2470 and 2630\,MeV/c$^2$.  The lowest of these states agrees in mass and width to the $d^*(2380)$ dibaryon.  Finer $W$ resolution and higher statistical precision is needed for confirmation however.

 A low mass enhancement in the $\pi^0\pi^0$ invariant mass distribution at low $W$ agrees with the ABC effect and adds credence to the $d^*(2380)$ contributing to the excitation spectrum.  With knowledge of the $d^*(2380)$ angular distribution, fitting to this spectrum enables an extraction of the $\gamma d \rightarrow d^*(2380)\rightarrow \pi^0\pi^0 d$ cross section of  $(11.3 \pm 3.2_\mathrm{stat} \pm 2.7_\mathrm{sys})$\,nb.

A peak in the $\pi^0 d$ invariant mass at 2114\,MeV/c$^2$ supports the case of the isovector dibaryon reported by the ELPH collaboration.  A striking feature of this new data is the narrow width of less than 20\,MeV/c$^2$, conflicting with the 91\,MeV/c$^2$ width reported at ELPH.  The differential cross section with respect to $W$ is described well by the proposed 2630\,MeV/$c^2$ dibaryon and supports the sequential decay of an isoscalar to an isovector dibaryon decay mechanism.  


\section*{Acknowledgements}

We would like to thank the staff and shift-students of the ELSA accelerator for providing an excellent beam.  We thank Alexander Fix,  Mikhail Bashkanov, Daniel Watts and  Eulogio Oset for insightful comments.

This work is supported by SFB/TR-16, DFG project numbers 388979758 and 405882627 and the Third Scientific Committee of the INFN.  This publication is part of a project that has received funding from the European Union’s Horizon 2020 research and innovation programme under grant agreement STRONG – 2020 - No.~824093.
P.~L.~Cole gratefully acknowledges the support from the U.S. National Science Foundation (NSF-PHY-1307340, NSF-PHY-1615146, and NSF-PHY-2012826) and the Fulbright U.S. Scholar Program (2014/2015).

\section{Erratum}

We would like to thank Alexander Fix for bringing the following small corrections to our attention.

\subsection{Normalisation of the  $\pi^0 d$ invariant mass distributions}

The $\pi^0 d$ invariant mass distributions in Fig.~8(a) and 9(a,c,e) include two values for every $\gamma d \rightarrow \pi^0\pi^0 d$ reaction due to the two possible $\pi^0 d$ combinations.  Consequently, the integrals when multiplied by the mass intervals give a differential cross section twice that of the differential cross section versus $W$.   Shown in Fig.~\ref{fig:pidmass1}(a) and Fig.~\ref{fig:pidmass} (left panels), this has now been corrected to avoid any confusion by scaling by a factor of 1/2.  

\subsection{A minor correction to the  $\pi^0 \pi^0$ invariant mass distributions}

From measurements of the incoming beam energy and deuteron momentum in the final state, the missing mass recoiling from the deuteron was used to determine the $\pi^0\pi^0$ invariant mass per event.  The perceived advantage of  this measurement instead of determining it from the two $\pi^0$ four-momenta was that the mass resolution is slightly better.  It has been  realised however that this introduces a small error, as the missing mass is occasionally  below the nominal two $\pi^0$ rest mass (approximately 270\,MeV/c$^2$) due to experimental resolution.  This is not the case however when determining the invariant mass from the two $\pi^0\rightarrow \gamma\gamma$ decays, as the two four-momentum vectors of the $\pi^0$ have their masses set to the nominal $\pi^0$ rest mass.  
This had the effect of slightly lowering the $\pi^0\pi^0$ invariant mass distributions, particularly close to threshold.  The $\pi^0\pi^0$ invariant mass is now determined from the two four-momenta of the identified $\pi^0$ to avoid this error (see Figs.~\ref{fig:pidmass1}(b), \ref{fig:pidmass} (right panels) and \ref{fig:dalitz}).

The main conclusions of the original paper remain the same, with one exception.  It was suggested that a low mass enhancement was evident in the $\pi^0\pi^0$ invariant mass close to threshold ($W = $ 2270 to 2441\,MeV) and that this may be the ABC effect which has been attributed to the $d^*(2380)$.  This reanalysis demonstrates that this was likely to be an artefact of the method of extraction of the invariant mass and is no longer as prominent in Fig.~\ref{fig:pidmass1}.  

\begin{figure} [h]
	\centering
	\vspace*{0cm}
	\resizebox{\columnwidth}{!}{%
		\includegraphics{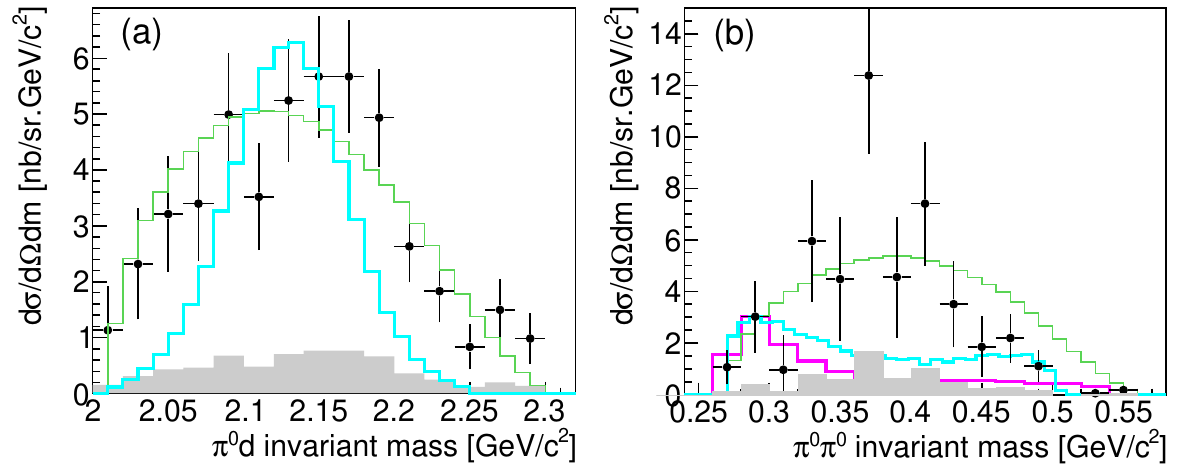}
	}
	\caption{Differential cross section versus the invariant mass of the (a) $\pi^{0}d$ and (b) $\pi^0\pi^0$ system for \dangle{} $>0.8$ and $W$ from  2270 to 2441\,MeV.  The measured data are the black data points and the systematic uncertainties are the grey bars on the absisca.  The green line is the phase space distribution with an integral equal to the measured data.  The $\pi^0\pi^0$ invariant mass includes the distribution from the ABC effect (magenta line),  with a scale fixed by the second data point at 0.29\,GeV/c$^{2}$.  This is a correction to Fig.~8.}		
	\label{fig:pidmass1}
\end{figure}

\begin{figure} [h]
	\centering
	\vspace*{0cm}
	\resizebox{\columnwidth}{!}{%
		\includegraphics{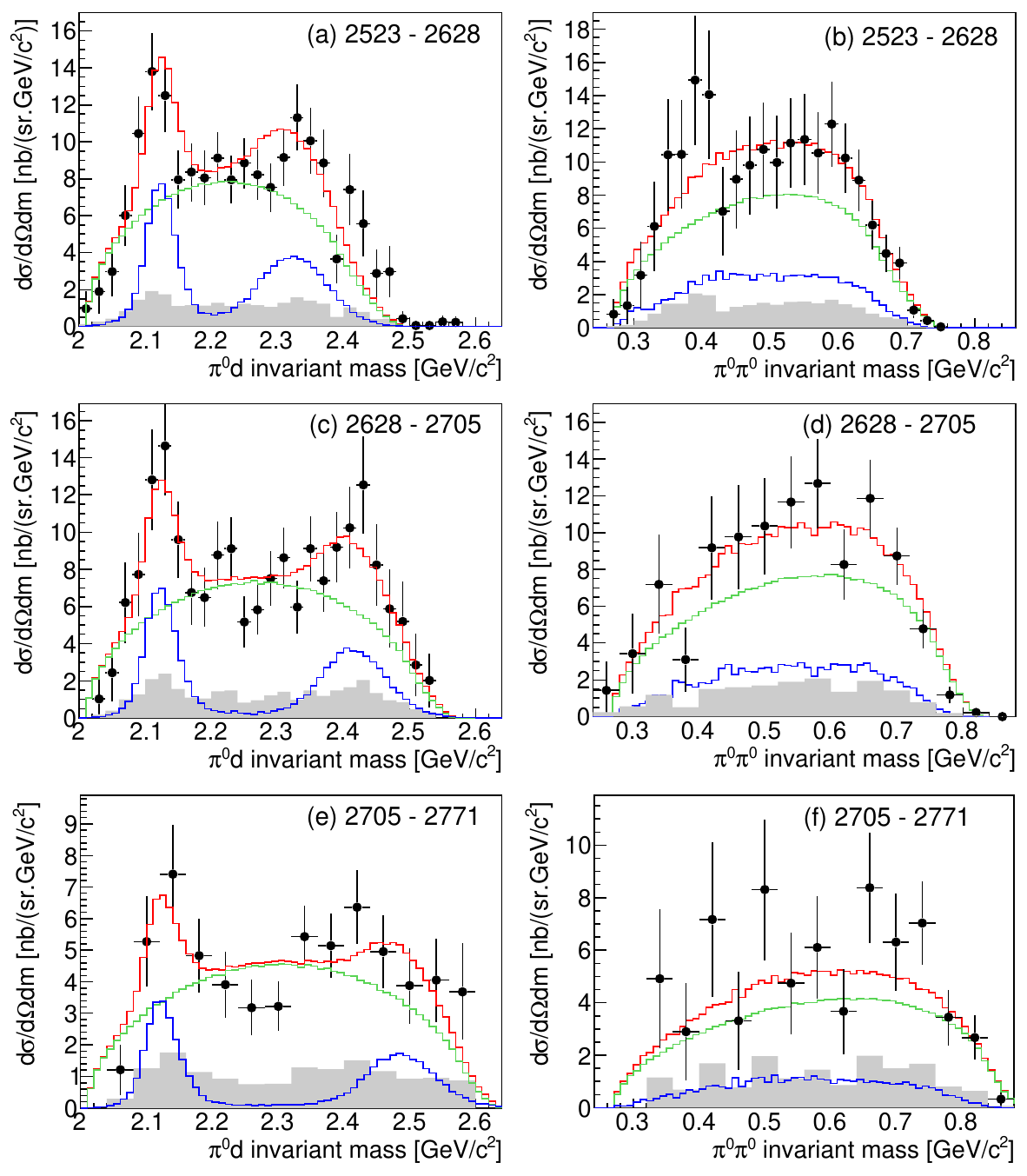}
	}
	\caption{Differential cross section versus the invariant mass of the (left panels) $\pi^{0}d$ and (right panels) $\pi^0\pi^0$ system for \dangle{} $>0.8$ and $W$ intervals labelled inset (corresponding to four tagger channels each).  The measured data are the black data points and the systematic uncertainties are the grey bars on the absisca.   The fitted distribution to the $\pi^0d$ invariant mass (red line) is comprised of phase space (green line) and proposed sequential dibaryon decay (blue line) contributions.  This is a correction to Fig. 9.}		
	\label{fig:pidmass}
\end{figure}

\begin{figure} [h]
	\centering
	\includegraphics[trim={0cm 0cm 0cm 0cm},clip,width=0.7\columnwidth]{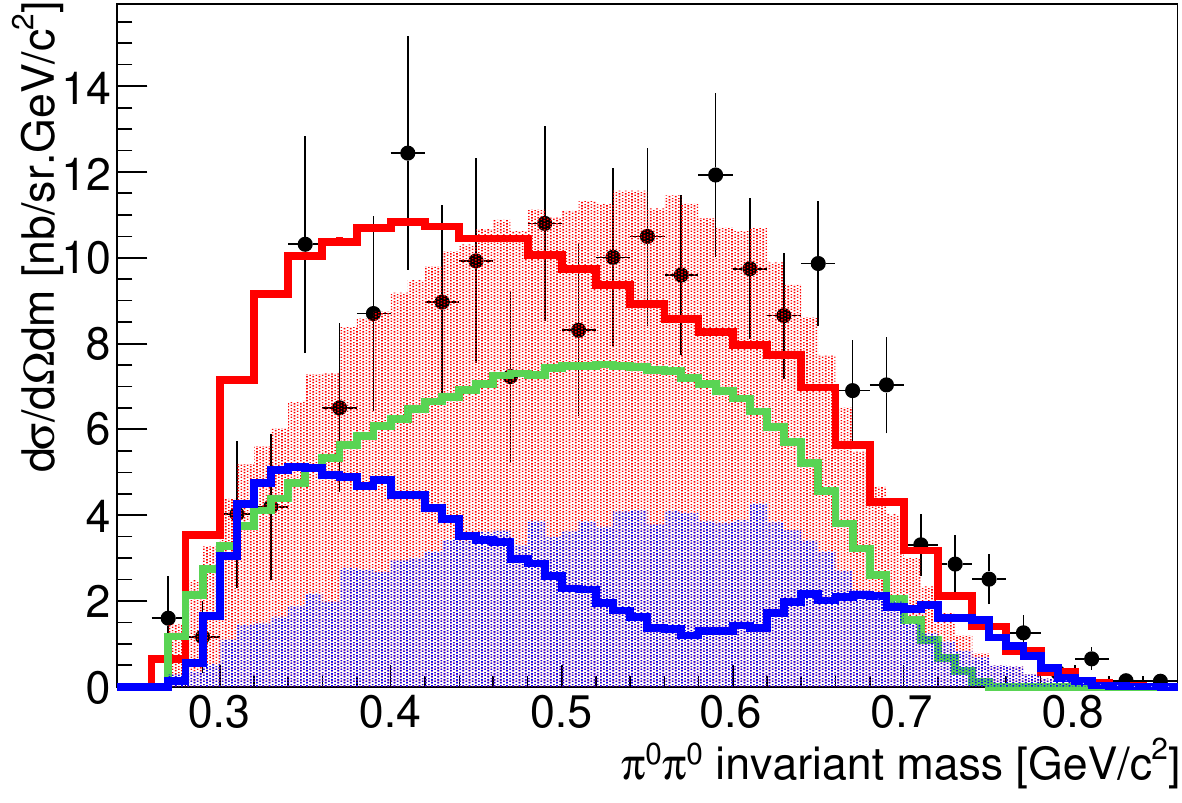}
	\caption{$\pi^0\pi^0$ invariant mass distribution for 2523 $< W < 2738$\,MeV (over the proposed sequential dibaryon decay) .  The shaded red distribution is comprised of the phase space (thick green line) and two sequential dibaryon decays both in $s$-wave (shaded blue area).
		The thick red line is comprised of the phase space and the 
		two sequential dibaryon decays both in $p$-wave (thick blue line).  This is a correction to Fig.~10.}		
	\label{fig:dalitz}
\end{figure}

\bibliographystyle{unsrt}

\begin{thebibliography}{}

\end{thebibliography}


\begin{thebibliography}{10}
	
	\bibitem{dyson64}
	F.~J. Dyson and N.-H. Xuong.
	\newblock {\em Phys. Rev. Lett.}, 13(26):815, 1964.
	
	\bibitem{clement17}
	H.~Clement.
	\newblock {\em Prog. Part. Nucl. Phys.}, 93:195, 2017.
	
	\bibitem{locher86}
	M.~P.~Locher, M.~E.~Sainio and A.~Svarc,
	\newblock {\em Adv. Nucl. Phys.}, 17:47, 1986.
	
	\bibitem{strakovsky91}
	I.~I.~Strakovsky,
	\newblock {\em Fiz. Elem. Chast. At. Yadra}, 22:615, 1991.
	
	\bibitem{yao64}
	Tsu Yao.
	\newblock {\em Phys. Rev. B}, 134:454, 1964.
	
	\bibitem{barnir73}
	I.~Bar-Nir.
	\newblock {\em Nucl. Phys. B}, 54:17, 1973.
	
	\bibitem{adlarson11}
	P.~Adlarson et~al.
	\newblock {\em Phys. Rev. Lett.}, 106:242302, 2011.
	
	\bibitem{bashkanov09}
	M.~Bashkanov.
	\newblock {\em Phys. Rev. Lett.}, 102:052301, 2009.
	
	\bibitem{clement21}
	H. Clement and T. Skorodko.
	\newblock {\em Chin. Phys. C}, 45:022001, 2021.
	
	\bibitem{adlarson13}
	P.~Adlarson et~al.
	\newblock {\em Phys. Lett. B}, 721:229, 2013.
	
	\bibitem{adlarson14}
	P.~Adlarson et~al.
	\newblock {\em Phys. Rev. Lett.}, 112:202301, 2014.
	
	\bibitem{adlarson14PRC}
	P.~Adlarson et~al.
	\newblock {\em Phys. Rev. C}, 90:035204, 2014.
	
	\bibitem{bashkanov19}
	M.~Bashkanov, S.~Kay, D.~P.~Watts et~al.
	\newblock {\em Phys. Lett. B}, 789:7, 2019.
	
	\bibitem{bashkanov20Pgamma}
	M.~Bashkanov, D.~P.~Watts, S.~J.~D.~Kay et~al.
	\newblock {\em Phys. Rev. Lett.}, 124:132001, 2020.
	
	\bibitem{adlarson13PRC}
	P.~Adlarson et~al.
	\newblock {\em Phys Rev. C}, 88:055208, 2013.
	
	\bibitem{adlarson15PLB}
	P.~Adlarson et~al.
	\newblock {\em Phys. Lett. B}, 743:325, 2015.
	
	\bibitem{goldman89}
	T.~Goldman, K.~Maltman, G.~J. Stephenson, K.~E. Schmidt, and Fan Wang.
	\newblock {\em Phys Rev. C}, 39(5):1889, 1989.
	
	\bibitem{gal14}
	A.~Gal and H.~Garcilazo.
	\newblock {\em Nuc. Phys. A}, 928:73, 2014.
	
	\bibitem{gal13}
	A.~Gal and H.~Garcilazo.
	\newblock {\em Phys. Rev. Lett.}, 111:172301, 2013.
	
		\bibitem{booth60}
	N.~E.~Booth, A.~Abashian, and K.~M.~Crowe.
	\newblock {\em Phys. Rev. Lett.}, 5:258, 1960.
	
	\bibitem{booth61}
	N.~E.~Booth, A.~Abashian, and K.~M.~Crowe.
	\newblock {\em Phys. Rev. Lett.}, 7:35, 1961.
	
	\bibitem{booth63}
	N.~E.~Booth, A.~Abashian, and K.~M.~Crowe.
	\newblock {\em Phys. Rev. C}, 132: 2296 1963.
	
%
%
%
	
	\bibitem{oh97}
	C.~H.~Oh, R.~A.~Arndt, I.~I. Strakovsky, and R.~L. Workman.
	\newblock{\em Phys. Rev. C}, 56:635, 1997.
	 
	\bibitem{arndt68} 
	R.~A.~Arndt.
	 \newblock{\em Phys. Rev.}, 165:1834, 1968.
	 
	 	\bibitem{bhandari81}
	 R.~Bhandari, R.~A.~Arndt, L.~D.~Roper and B.~J.~VerWest.
	  \newblock{\em Phys. Rev. Lett.}, 46:1111, 1981.
	
	
	\bibitem{molina21}
	R.~Molina, N.~Ikeno, and E.~Oset.
	\newblock arXiv:2102.05575v3 [nucl-th], 2021.
	
	\bibitem{ikeno21}
	N.~Ikeno, R.~Molina, and E.~Oset.
	\newblock {\em Phys Rev. C}, 104:014614, 2021.
	
	\bibitem{bashkanov21}
	M.~Bashkanov, H.~Clement and T.~Skorodko.
	\newblock arXiv:2106.00494v2 [nucl-th], 2021.
	
	\bibitem{dong16}
	P.~Shen Y.~Dong, F.~Huang and Z.~Zhang.
	\newblock {\em Phys Rev. C}, 94:014003, 2016.
	
	\bibitem{vidana18}
	I.~Vida\~{n}a, M.~Bashkanov, D.~P. Watts, and A.~Pastore.
	\newblock {\em Phys. Lett. B}, 781:112, 2018.
	
	\bibitem{bashkanov20}
	M.~Bashkanov and D.~P. Watts.
	\newblock {\em J. Phys. G: Nucl. Part. Phys}, 47:03LT01, 2020.
	
	\bibitem{ishikawa17}
	T.~Ishikawa et~al.
	\newblock {\em Phys. Lett. B}, 772:398, 2017.
	
	\bibitem{fix05}
	A.~Fix and H.~Arenh\"ovel.
	\newblock {\em Eur. Phys. J. A}, 25:115, 2005.
	
	\bibitem{egorov15}
	M.~Egorov and A.~Fix.
	\newblock {\em Nucl. Phys. A}, 933:104, 2015.
	
	\bibitem{guenther17}
	M.~S. Guenther.
	\newblock {\em PoS(Hadron2017)}, 210:051, 2018.
	
	\bibitem{ishikawa19}
	T.~Ishikawa et~al.
	\newblock {\em Phys. Lett. B}, 789:413, 2019.
	
	\bibitem{technicalpaper}
	S~Alef et~al.
	\newblock {\em Eur. Phys. J. A}, 56:104, 2020.
	
	\bibitem{hillert06}
	W.~Hillert.
	\newblock {\em Eur. Phys. J. A}, 28:139, 2006.
	
	\bibitem{hillert17}
	W.~Hillert et~al.
	\newblock {\em EPJ Web Conf.}, 134:05002, 2017.
	
		\bibitem{klambdapaper}
	S.~Alef et~al.
	\newblock {\em Eur. Phys. J. A}, 57:80, 2021.
	

	\bibitem{kashevarov12}
	V. L Kashevarov et~al.
	\newblock {\em Phys. Rev. C}, 85:064610, 2012.
	
	\bibitem{risser73}
	T.~Risser and M.~D. Shuster.
	\newblock {\em Phys. Lett. B}, 43:68, 1973.
	
	\bibitem{bar75}
	T.~Risser I.~Bar-Nir and M.~D. Shuster.
	\newblock {\em Nucl. Phys. B}, 87(109), 1975.
	
	\bibitem{alvarez98}
	E.~Oset L.~Alvarez-Ruso and E.~Hernandez.
	\newblock {\em Nucl. Phys. A}, 633:519, 1998.
	
	\bibitem{alvarez99}
	L.~Alvarez-Ruso.
	\newblock {\em Phys. Lett. B}, 452:207, 1999.
	
	
		\bibitem{adlarson12}
	P.~Adlarson et~al.
	\newblock {\em Phys. Rev. C}, 86:032201, 2012.
	
	\bibitem{dong19}
	Yubing Dong, Pengnian Shen, and Zongye Zhang
	\newblock {\em Int. J. Mod. Phys. A}, 34:1950100, 2019.
	
	
	\bibitem{pdg20}
	P.~A.~Zyla et~al. (Particle Data~Group).
	\newblock {\em Prog. Theor. Exp. Phys.}, 083C01, 2020.
	
\end{thebibliography}
\end{document}